\documentclass[]{article}
\usepackage[utf8]{inputenc}
\usepackage[margin=1in]{geometry}
\usepackage[round]{natbib}
\usepackage{dialogue}
\usepackage{fancyhdr}
\usepackage{amssymb}
\usepackage{amsfonts}
\usepackage{amsmath}  
\usepackage{setspace}

\title{When is it permissible for artificial intelligence to lie?:\\ A trust-based approach}

\author{Tae Wan Kim\thanks{Associate Professor of Business Ethics, twkim@andrew.cmu.edu, Carnegie Mellon University}, Tong (Joy) Lu\thanks{Assistant Professor of Marketing, Carnegie Mellon University}, Kyusong Lee\thanks{Co-founder, SOCO.AI},\\ Zhaoqi Cheng\thanks{PHD student, Carnegie Mellon University}, Yanhan Tang\thanks{PHD student, Carnegie Mellon University} and John Hooker\thanks{Professor of Operations Research, T. Jerome Holleran Professor of Business Ethics and Social Responsibility, Carnegie Mellon University}}

\begin{document}

\maketitle

\begin{abstract}
\noindent Conversational Artificial Intelligence (AI) used in industry settings can be trained to closely mimic human behaviors, including lying and deception. However, lying is often a necessary part of negotiation. To address this, we develop a normative framework for when it is ethical or unethical for a conversational AI to lie to humans, based on whether there is what we call ``invitation of trust'' in a particular scenario. Importantly, cultural norms play an important role in determining whether there is invitation of trust across negotiation settings, and thus an AI trained in one culture may not be generalizable to others. Moreover, individuals may have different expectations regarding the invitation of trust and propensity to lie for human vs. AI negotiators, and these expectations may vary across cultures as well. Finally, we outline how a conversational chatbot can be trained to negotiate ethically by applying autoregressive models to large dialog and negotiations datasets.
\end{abstract}


\noindent As businesses increasingly utilize artificial intelligence (AI) to make important decisions for humans, societal worries about the compatibility of AI and human values grow \citep{martin2019business, martin2019designing, martin2019ethical, hooker2018, HK, 8511831, Kim2019, kim2019artificial}. In response, researchers have examined how to make AI learn ethical principles as well as calculative and strategic intelligence \citep{bhargava2017autonomous, kim2018mimetic, kim2020taking}. Such attempts are lumped under the broader term ``value alignment''\citep{russell2015research}. In this project, we attempt to ground conversational AI's behavior in ethical values. 

\section{The apparent problem}

\noindent End-to-end dialogue and conversational agents trained on large human datasets have been shown to mimic both ethical and unethical human behaviors. For instance, goal-oriented dialogue agents can learn deceptive and misleading behaviors from human negotiators. In fact, from Facebook's end-to-end dialogue system, \citeauthor{DBLP:journals/corr/LewisYDPB17} (\citeyear{DBLP:journals/corr/LewisYDPB17}) reported:

\begin{quote}
    ``Analysing the performance of our [dialogue] agents, we find evidence of sophisticated negotiation strategies. For example, we find instances of the model feigning interest in a valueless issue, so that it can later `compromise' by conceding it. Deceit is a complex skill that requires hypothesizing the other agent's beliefs, and is learnt relatively late in child development. {\em Our agents have learnt to deceive without any explicit human design simply by trying to achieve their goals}''\citep[Italics added]{DBLP:journals/corr/LewisYDPB17}
\end{quote}

\noindent As more companies are using conversational AI, it is worrisome that AI can potentially mislead, deceive, and/or manipulate humans. In this article, we explore how to ground conversational AI to behave ethically. 

We develop a normative framework of when it is unethical for AI to lie to humans, using cases of a negotiation, which we find distinctively useful to clarify the ethics of AI lies. We also explain why the ethical status of conversational AI in part depends upon empirical norms (e.g., cultural expectations). By doing so, we discuss the possibility that a particular kind of AI lie that is unethical in a certain culture is not necessarily unethical in a different culture. Finally, we discuss how to develop a conversational AI that is capable of identifying when it is permissible or not to lie to humans. 

For the question of whether AI {\it can} lie, please see the Appendix. Here we assume that the AI agent can lie or at least utter false statements with a goal to deceive.

\section{The ethics of lies in negotiation}
One might think that conversational AI must be regulated to {\em never} utter false statements (or lie) to humans. But, the ethics of lying in negotiation is more complicated than it appears. Lying in negotiation is not necessarily unethical or illegal under some circumstances, and such permissible lies play an essential economic role in an efficient negotiation, benefiting both parties \citep{shell1991legal, strudler1995ethics, strudler2005deception}. Consider the simple scenario of a used car negotiation: 

\begin{dialogue}
\speak{consumer} Hi, I'm interested in used cars.
\speak{dealer} Welcome. I'm more than willing to introduce you to our certified pre-owned cars.\\
\noindent \ldots{} 
\speak{Consumer} I'm interested in this car. Can we talk about price?
\speak{Dealer} Absolutely. I don't know your budget, but I can tell you this: You can't buy this car for less than \$25,000 in this area. [Dealer is lying] But it's the end of a month, and I need to sell this car as soon as possible. My offer is \$24,500.  
\speak{Consumer} Well, my budget is \$20,000. [Consumer is lying] Is there any way that I can buy the car for around \$20,000?  \\
\noindent \ldots{}  
\end{dialogue}

During the negotiation, both parties lied to each other about their reservation prices. Interestingly, both lied {\it not} unethically (and not illegally). Why? A short explanation is that no one was duped or harmed. But to explain more precisely, we need an excursion into normative ethics theory.

First, we realize that there are multiple ways to define lying \citep{sep-lying-definition}, and a single definition cannot cover all contexts. For our context---Human-to-AI--, we find the following to be apt:

\begin{quote}
    Let ``{\em x} lies to {\em y}'' := There is a proposition {\em P} such that (i) {\em x} believes that {\em P} is not true (or does not believe that {\em P} is true)  and (ii) {\em x} asserts {\em P} to {\em y} \citep{chisholm1977intent}.\footnote{Note that, by this definition, a lie is not necessarily intended to deceive.  ``Lie'' is often interpreted as intentional deception. For more about this issue, see the Appendix.}
\end{quote}

Now let us explain why the dealer's lie was not unethical, drawing upon a trust-based account of lying \citep{fried1978right,fried2015contract,strudler2005deception, williams2002truth}. Contextually, neither of the two parties implicitly assured or invited the other to trust that they would tell true statements about the reservation price. When the dealer asserted that the bottom price was \$24,500, the consumer contextually construed it as signalling, for instance, ``This is my {\it initial} offering. I am open to further negotiating with you.'' The dealer, as well, did not take the consumer's statement at face value. There was no assurance or invitation of trust. Therefore, there cannot be breach of a trust in the dialogue. If there is breach of a trust involved in a lie, that is an unethical case of lying. If not, it is not an unethical case. 

As a legal matter, ``reliance''---one of the conditions of Common Law Fraud--- plays the role of ``trust.'' If there is no invitation to trust, there is no reliance. A consumer reasonably does not rely on a car dealer's statements about reservation price, so the statements do not establish fraud. In this paper, we focus on ethics. Historically, courts respect moral principles as well as the black letter of negotiation laws. As G. Richard Shell (\citeyear{shell1991legal}) pointed out, ``Unethical bargaining practices are, as often as not, illegal or become so after they are brought to light. The law simply expands to include them, definitions notwithstanding.'' 

We do not deny that there can be counter-examples to the trust-based approach. But it is a well-accepted theory in the ethics of negotiation. Furthermore, the trust-based account can be further explained and normatively justified by more generic moral principles such as universalization/generalization, autonomy and utility-maximization \citep{Hooker2018toward}. To use a Kantian term, lying about a reservation price in negotiation is generalizable/universalizable, because any parties who lie about reservation prices in a competitive negotiation can consistently achieve their goals. In fact, lying about reservation prices is a generalized practice in the commercial domain. Second, lying about a reservation price does not involve a breach of autonomy, because there is an explicit or implicit consent relation in which statements about reservation prices are intended or foreseen to be false or misleading. Parties are aware of the risk. Finally, deception about reservation prices tends to enhance the total utility of both parties. So, lying about reservation prices is not unethical according to any of the three generic ethical principles (generalization, autonomy, and utility). In contrast, lying about safety issues does not meet generalization or autonomy requirements, although it can sometimes enhance utility.

To clarify the role of a trust in the ethics of lying in negotiation, consider the following where there is breach of a trust:

\begin{dialogue}
\speak{consumer} Hi, I'm interested in a pre-owned/used car.
\speak{Car dealer} Welcome. I'm more than willing to introduce you to our pre-certified cars.
\noindent \ldots{} several questions and answers about technical functions \ldots{}
\speak{Consumer} How many airbags does this car have? Are they safe enough?
\speak{Car dealer} This car has 6 airbags originally certified by NHTSA and recently by our professional mechanics. [The dealer is lying.]
\speak{Consumer} That's great. 
\end{dialogue}

In this conversation, the consumer is conventionally invited to trust the dealer's statement about safety. The dealer invites the consumer to trust his propositions about safety, by asserting that the airbags are certified. But the dealer is, in fact, lying. The airbags were not recently certified by professional mechanics. Thus, the dealer breaches the consumer's trust that he invited her to have toward him. A lie that involves a breach of trust is unethical. Similarly, a promise or a contract must be kept because it is wrong to breach the trust that the promisor has invited from the promisee by performing the speech act, ``I promise you'' \citep{fried1978right}.   Thus, when the dealer lies about safety, that is an unethical lie. In contrast, when the dealer lies about reservation prices, there is no breach of trust, so such a lie is ethically permissible. 

In a trust relationship, {\em A} expects {\em B} to act not simply as {\em A} assumes {\em B} will, but as {\em B} should. So, {\em A} has {\it normative}, more than predictive, expectations. So, when {\em A} is invited to trust {\em B} to do X, {\em A} holds \textit{B} responsible for doing X. Thus, \textit{B} is trustworthy to the extent that \textit{B} fulfills the normative expectations. 

To summarize, we offer an account of when it is unethical and not unethical to lie, as follows:

\begin{quote}
    Let ``{\em x} lies to {\em y} unethically'' := There is a proposition {\em P} such that (i) {\em x} believes that {\em P} is not true  (or does not believe that {\em P} is true) and (ii) {\em x} asserts {\em P} to {\em y}, and (iii) there is a breach of a trust (i.e., Given cultural conventions and societal expectations, it is rational for {\em y} to believe that {\em x} (implicitly) invites or assures {\em y} to trust {\em x} not to lie to {\em y}, but {\em x} lies to {\em }y).
\end{quote}

\begin{quote}
    Let ``{\em x} lies to {\em y} not unethically'' := There is a proposition {\em P} such that (i) {\em x} believes that {\em P} is not true  (or does not believe that {\em P} is true)  and (ii) {\em x} asserts {\em P} to {\em y}, and (iii) there is no invitation/breach of a trust (i.e., Given cultural conventions and societal expectations, it is rational for {\em y} to believe that {\em x} does not invite or assure {\em y} to trust {\em x} not to lie to {\em y}). 
\end{quote}

Invitation of trust is not a purely subjective attitude. It is in part an objective, conventional state of affairs \citep{grice1991studies}. Even if you want to invite the other party's trust when you negotiate a price, it is, in most circumstances, hard to do so, given the cultural influence of social expectations.  Consider this:

\begin{dialogue}
\speak{Consumer} I like this car. Can we talk about price?
\speak{Car dealer} Well, you can't buy this car for less than \$32K in this area. My offer is \$32K. That's the lowest price in this dealership. [The dealer is not lying.]
\speak{Consumer} That's a lot higher than my budget. [She believes that the dealer's price is literally an offering price and the dealer is willing to go below the price.]
\speak{Car dealer} As I said, it's literally impossible to go below 32. [The dealer is not lying.]
\speak{Consumer} ..... How about 26K?
\speak{Car dealer} Again, my bottom line is 32K. [The dealer is not lying.]
\speak{Consumer} .....  Okay, good bye. [The consumer feels disappointed and even slighted.]
\end{dialogue}

Negotiators are conventionally expected to discredit statements about reservation prices, at least in North America. A negotiator such as the dealer above, who does not lie, will likely put the whole process of negotiation into jeopardy, unless he finds a reliable way to communicate his truthfulness to the other party. If the dealer's motivation was not to behave unethically, he did not fail. But he failed to achieve economic benefit for the other party as well as his. What he did not know was that even if he had lied about reservation prices, he would not have acted unethically. At the same time, both parties would have gained economic benefits. This shows that developing a {\it moralistic} chatbot that never lies can be inferior to a chatbot that can lie ethically. To achieve both ethical and economic benefits at the same time, negotiation-bots must be trained to identify the circumstances where it is ethically permissible or impermissible to lie. In the last section, we briefly discuss how to train a negotiation chatbot to identify these circumstances in the domain of a car negotiation.

\section{The ethics of conversational AI lies}

Now imagine that you want to buy a car from a company that sells cars through an online website that allows for price negotiation and has an AI chatbot dealer that is ready to negotiate with you. Consider the following scenario: 

\begin{dialogue}
\speak{Consumer} How many airbags does this car have?
\speak{AI dealer} It has 5 airbags. 
\speak{Consumer} Can we talk about price?
\speak{AI dealer} Well, you cannot buy this car for less than \$65,000 in this area. My offer is \$63,500. That's the lowest price in this dealership. [AI is lying.] 
\end{dialogue}

\noindent The negotiation chatbot's statement,``Well, you cannot buy this car for less than \$65,000 in this area. My offer is \$63,500K. That's the lowest price in this dealership.'' is not true. So, the statement is a lie. 

Asserting whether the statement is ethically permissible or not depends upon whether there was an invitation to trust. It is in part an empirical matter whether consumers are conventionally invited to trust that negotiation chabots are asserting true statements. Unlike in human-to-human interaction, it is possible that humans trust negotiation bots no matter what subject matter (e.g., safety, price) is discussed. Another issue with chatbots is that they do not provide nonverbal signals, which are often very important in negotiation and vital clues to whether the speaker is inviting trust.  For example, in Western cultures, not looking someone in the eye is a warning to distrust the speaker (very different than in parts of Asia, where it may be a way of showing respect).

\subsection{Economic model}
To further clarify the ethics of AI lies, we mathematically express our account by the idiom of an economic model, ``signal jamming'' \citep{stein1989efficient,holmstrom1999managerial}. The goal of the chatbot on behalf of a business firm is to maximize its utility, a problem that can be formulated as follows:

\[
\max_b \; \theta E(x|y) + (1-\theta)E(x) - C(b)
\]

where $b$ is the amount of lying, $x$ is the true value (e.g., price) of the conversational AI to a consumer, and $y$ is the value (e.g., initial price) uttered by the AI. $E(x|y)$ is what the consumer expects from the AI, given the value $y$ uttered by the AI. $E(x)$ is what the consumer expects on the basis of other sources, such as existing information about how AI generally behaves, etc. The multiplier $\theta$ is an estimate of how much the consumer's overall estimation depends on the AI's utterances, where $0\leq\theta\leq 1$. $C(b)$ is the expected cost to the AI on behalf of the company of an amount $b$ of lying, perhaps due to loss of overall reputation.\footnote{This section benefits from \citeauthor{kim2020yuji} (\citeyear{kim2020yuji})'s analysis of signal jamming for accounting fairness.}

Suppose that in equilibrium, consumers know that a certain amount $\hat{b}$ of lying occurs, for instance, about reservation price. Or $\hat{b}$ is the amount of lying that a consumers expects. This means that
\[
E(x|y) = y - \hat{b} 
\]

If we make the model as an optimization problem, it now becomes
\[
\max_b \; \theta (E(x)+b-\hat{b}) + (1-\theta)E(x) - C(b)
\]
Since $\hat{b}$ is a constant, this implies that the optimal value $b^*$ of $b$ satisfies
\[
C'(b^*)=\theta
\]
Thus, the AI on behalf of the company is incentivized to lie in negotiation or advertising until the marginal cost of lying is $\theta$; that is, until adding one unit of value by lying is offset by a cost of $\theta$. If the consumer relies solely on the AI's utterances, then $\theta=1$, and the AI adds lying until a point of vanishing returns; that is, until the value of any additional lying incurs an equal expected cost. Typically, $C'(b)>0$ when $b>0$, which means that lying exists in equilibrium. Finally, since consumers know how much lying occurs in equilibrium, $b^*=\hat{b}$. 

When does $b^*=\hat{b}$ occur? In our context, it occurs when the AI invites the consumer to not trust its utterance. When it happens, the AI lying is not unethical. The basic intuition of signal-jamming in our context is that if the AI lies without inviting consumers to trust it, consumers will accordingly discredit the AI's utterances to the extent that AI lies to consumers. Thus, consumers will not be affected by the AI's lies. Furthermore, in that situation, a moralistic AI that never lies will put both the consumers and the corporation that uses the AI into a disadvantaged position. 

\subsection{How can AI identify whether $b^*=\hat{b}$?}
A real question is how AI can identify whether $b^*=\hat{b}$ in a specific domain or context. The trust-based framework we offered above is useful here. Whether there is an invitation of trust between AI and humans can be empirically surveyed with domain-specific scenarios.

Here are examples of how to collect datasets. The most intuitive way to collect conversational data for our purposes is to let a machine learn from large amounts of real human-to-AI conversations, for instance in negotiation settings. It is possible to conduct an online negotiation between AI sellers and human buyers. Each group of participants might have have the goal to maximize profits. The best buyer and seller will be awarded at the end of the experiment. To collect more realistic training data for the chatbot, it is possible to use the ``Wizard-of-O'' (WOZ) data collection method \citep{okamoto2001wizard}: Subjects interact with a computer system that they believe to be an AI, but is actually being operated or partially operated by an unseen human being.

Alternatively, scenario dialogs in negotiation can be given to participants in controlled lab studies in order to (1) gauge participants' expectations regarding the trustworthiness and/or honesty of the negotiator in a particular setting, and (2) determine whether these expectations vary between human-to-human and human-to-AI interactions. Possible scenarios include price negotiation with AI, interaction with advertising and marketing text generated by AI, and customer service settings. For example, participants may be asked to imagine purchasing a used car and negotiating with a seller, who may be a human or an AI chatbot. To determine participants' expectations regarding whether or not the seller will lie to them, direct questions such as ``How much do you believe that the given price is their lowest possible offer'' may be asked, or indirect measures such as rating their surprise upon discovering that the seller had lied (e.g., finding out that another buyer had been given a lower offer). It is also important to test both scenarios where participants typically would expect an invitation of trust in a real-world human-to-human setting (e.g., regarding safety features such as airbags) and scenarios where the expectation is that one or both parties may lie (e.g., reservation prices).

In this paper, we focus on theory. In another paper, our team is currently doing survey projects to identify when humans believe that they are invited by AI to trust or not. We survey different industry-specific scenarios and how people living in different countries react differently.  

\section{Cultural factors}
In this section, we explain why it is permissible for AI to lie about price in the U.S., but impermissible in a Scandinavian country. For instance, the degree of invitation of trust in human-to-human price negotiation in a Scandinavian country is believed to be significantly higher than in the U.S. The same thing can happen in Human-to-AI interaction. Then, a conversational AI ethically optimized for U.S. consumers is not ethically ideal for consumers in Scandinavian countries. To avoid this problem, empirical research is needed to measure different degrees of invitation of trust in different cultures.

It must be noted that our approach is neither morally relativistic, nor absolutist. We use the same normative standard for the U.S. and the Scandinavian countries, so it is not moral relativism. Simultaneously, the approach is consistently sensitive to cultural norms. Hence, let use categorize ours as moral pluralism. 

The fundamental purpose of negotiation, worldwide, is to identify a mutually acceptable exchange.  However, the manner in which agreement is reached can vary significantly across cultures.  One must keep this in mind when evaluating the ethics of lying in negotiation. We provide only a brief overview of cultural factors here; a fuller account is given by \citeauthor{Hoo12} (\citeyear{Hoo12}). 

Nothing in our discussion is meant to suggest that one type of  culture is more ethical than another.  We only observe that cultures have different behavioral norms because people have developed different ways of living together.  We also recognize that not everyone within a given culture follows standard practices.  Cultural norms are to individual behavior as climate is to weather: the norms govern the overall functioning of the culture, while a great deal of individual variation can occur without undermining the basic mechanism.  

To simplify matters, let's suppose party A wishes to sell an item to party B, and they are negotiating a price.  Party A has a lowest price she is willing to accept, and party B a highest price he is willing to pay.  If the seller's reservation price is lower than the buyer's, a deal is possible in principle.  However, if the parties reveal their reservation prices to each other, then A will insist on asking B's reservation price, because she knows he is willing to pay a price that high, while B will insist on A's reservation price, because he knows she is willing to accept a price that low.  Negotiation therefore breaks down.  Somehow, the parties must find a way to agree on a price without revealing their reservation prices.

The typical escape from this impasse is to use culturally-specific signaling to convey a limited amount of information about each party's reservation price.  The simplest type of signal is an offer.  If B offers a price, A knows that he is willing to pay at least that much and makes a counteroffer on that basis.  When she makes the counteroffer, B knows that she is willing to accept at least that amount and revises his initial offer accordingly. Ideally, the offers will converge, but the process can stall, as when each party knows too little about the other to risk an offer that differs too much from the previous one.

To accelerate the process, cultures have developed additional signaling conventions.  The difference between one's current offer and the previous one may indicate how close he/she is to a reservation price.  Facial expressions and other body language, feigned emotions, and walking out of the shop can transmit additional information, but not too much.  This kind of back-and-forth may include statements that are false if interpreted literally.  Yet they are not taken at face value because the parties understand them as revealing information about mutually agreeable prices. The signaling conventions can be quite subtle and nearly impossible for outsiders to pick up on.  A full understanding may require that one grow up in the country and watch one's elders negotiate in many settings. 

Even this general pattern may break down in some cultures.  Strongly rule-oriented cultures, such as those of Scandinavia, may be based on a social contract that requires full disclosure of each party's preferences and situation from the outset.  The purpose of negotiation is then seen as achieving maximum benefit for the two parties, almost as though they were on the same side.  

In addition, Western cultures frequently use a low-context communication style in Edward Hall's sense (\citeyear{Hal76}), meaning that information tends to be transparently available.  The United States is a textbook example.  These cultures tend to prefer a fixed selling price that is \mbox{revealed} to the public, as on a price tag or marketing website.  People are willing to negotiate large transactions, such as an automobile purchase or a business deal, but generally dislike negotiation and prefer not to ``waste time haggling.''  Acceptable negotiation practices may vary from one industry or commercial setting to another rather than follow general cultural norms.  

At the other extreme, business people in relationship-oriented cultures may prefer to develop long-term mutual trust relationships before undertaking a joint venture.  In China, for example, this kind of relationship is known as {\em guanxi}.   Across-the-table negotiation in the usual sense is unnecessary, as the parties already understand each other's situation and wish to maintain a valuable relationship by making sure that both parties benefit from the venture.  On the other hand, negotiation with strangers in a street market follows the give-and-take pattern described above.  

An empirical study of negotiation behavior must take careful notice of the surrounding culture and the business context within that culture.  It must also recognize, as already noted, that online negotiation may differ from person-to-person negotiation due to its inability to take into account gestures and other body language, particularly in cultures where they are an essential part of communication.  

\section{How to train a chatbot to negotiate ethically}
In this section, we sketch our future plan to actually develop an ethical conversational AI.

\subsection{Data collection}
Datasets play a crucial role in modern AI. Many AI researchers call data the ``rocket fuel'' of AI. Collecting sufficient conversational data in negotiation will play a key role in successfully training a chatbot to negotiate ethically. In particular, high quality data is best for training data in data-driven approaches. However, it is expensive to collect high quality conversational data to achieve reasonable performance in a data-driven chatbot. Therefore, many commercial chatbots are built with rule-based approaches \citep{weizenbaum1966eliza}, which are relatively straightforward and can achieve good performance in specific target domains. However, writing rules for different scenarios is expensive and time consuming, and it is impossible to write rules for every possible scenario \citep{vinyals2015neural}. In order to train a chatbot to lie ethically with real AI, it is important to efficiently collect conversational data in negotiation. It is important to have enough high quality data to train the machine learning model. As we discussed, scenario-based dialog can be used to measure the degree of invitation of trust to detect how much a user will trust the chatbot's utterances. Finally, it will be critical to guard against teaching a bot unethical negotiation. Detecting the invitation of trust will be key in preventing unethical negotiation. When the assurance of trust is less than the threshold, the bot will always tell the truth in order to prevent unethical negotiation. Moreover, using reinforcement learning, a high penalty must be put on the reward function for unethical negotiation action. 

\subsection{Model training}
Our goal is to build a general dialog model to negotiate ethically using diverse datasets. Such a dialog system is much more complex than traditional dialog systems, such as slot-filling info giving systems or 1-turn question answering, because it requires the dialog model not only to understand the subtle semantics in user inputs, but also to generate natural and appropriate responses (some of which are ethically permissible lies). As a result, we will build upon the state-of-the-art end-to-end dialog models, which can in principle imitate any types of conversations, given enough training material.

We plan to create this model in 3 steps: pre-training, domain fine-tuning, and reinforcement learning.

\subsubsection{Pre-training}

The first step involves leveraging a huge amount of human-to-human conversational data to create pre-training models that learn the backbone of conversations, including word embeddings, sentence embeddings, discourse relationships, and typical conversational flow \citep{zhao2018unsupervised}. The reason for this is that it will be expensive to collect high-quality dialog data that exhibits the natural human  behavior of ethically permissible lies. Although it is plausible to collect high-quality data for our use on the scale of thousands, it is far less than the typical data size needed for large deep learning models, which are on the scale of millions \citep{zhao2018zero}. This is motivated by recent advances in pre-training on massive amounts of text data that have led to state-of-the-art results in a range of natural language processing (NLP) tasks  \citep{peters2018dissecting,devlin2018bert,radford2018improving} including natural language inference, question answering, and text classification. 

We will leverage state-of-the-art autoregressive models, including XLNet \citep{yang2019xlnet} and  GPT-2 \citep{radford2019language}, and use large dialog datasets (e.g.,  OpenSubtitle, Twitter, Reddit Conversations, etc.) for training \citep{serban2018survey}. The basic architecture will create an autoregressive language model that predicts a missing word given the context. This approach has been proven to create state-of-the-art language understanding systems  \citep{devlin2018bert}. The resulting models will be passed on to the second step for domain-specific fine-tuning.

\subsubsection{Supervised fine-tuning:}
In this step, the pre-trained models will be adapted to fulfill the end-task, a full-fledged dialog system \citep{mehri2019pretraining}. We will use the next utterance generation/retrieval as the objective functions. That is, given an arbitrary dialog context from the training data, i.e., a segment of an ongoing conversation, the model is required to predict the next correct response in context. This is essentially an imitation learning process where we create models to imitate how humans speak in our desired scenarios. Moreover, since we are using pre-trained models as a starting point, the convergence of the system will be faster than training from scratch, and end performance will be better. 

More concretely, we want to optimize the model parameters to maximize the conditional distribution $P\left(X \vert C\right)$, where X is the response and C is the dialog context. This can be done via gradient-based optimization techniques on neural network models. To prevent the model from overfitting to smaller domain-specific data, we will apply various fine-tuning techniques to prevent degenerate behavior such as catastrophic forgetting. In the end, this step will create an initial model that can converse with human users and achieve reasonable dialog performance. 

\subsubsection{Reinforcement learning}
In order to further improve the performance of the system, reinforcement learning may be applied \citep{zhao2019rethinking}. Unlike supervised fine-tuning, the reinforcement learning steps encourage the model to discover its own optimal decision-making policy in order to maximize a utility function, which we can define. It is useful in our setting because ethically permissible lies may be sparse events in real human-to-human dialog; therefore, the model may not learn them well enough since they are not a major factor contributing to the conditional likelihood. However, if the utility function is directly related to whether or not the model is able to learn ethically justified lies, then the model can create much better representations in order to master the ability to correctly produce permissible lies. We will apply policy gradient reinforcement learning methods, e.g., REINFORCE, A2C etc., to further tune the models from step 2. Ultimately, the resulting model should have the ability to produce responses that are superior in terms of appropriateness in the context of natural language and ethics.

\section{Concluding remarks}

The increasing use of conversational AI in negotiations across the globe raises the concerns that AI may lie and potentially deceive humans. However, in certain scenarios, AI lies may not necessarily be unethical. In this paper, we developed a normative framework for determining when AI lying is unethical in the context of negotiation. We also discussed how and why empirical norms, specifically, cultural expectations, may affect the ethical status of conversational AI. To illustrate how to build an ethical conversational AI, we elaborated on steps and procedures to train
a general dialog model. This paper addresses the theory of ethical conversational AI; empirical studies on conversational AI ethics under different cultural norms are left for future research.

Although our discussions focus on AI ethics in negotiation, our proposed framework and conclusions may extend to other AI contexts where the communication and the invitation of trust between human and AI are involved for decision making; examples include job interviews in e-recruitment, clinical interviews in tele-medicine, and virtual agents in e-commerce or online real-estate business, as often documented in literature \citep{van2019marketing,larosa2018impacts,honarvar2010towards}.
As the magnitude of human trust in AI may vary across these different contexts and scenarios, our proposed normative framework may help business people and service providers to understand the ethical issues in each specific domain, avoid unintended unethical behaviors, and further design AI systems that make ethical decisions.

There are several potential future directions that may build on and generalize the results of our paper. First, it might be worthwhile to investigate empirically how specific cultures respond to AI lies and to which extent the underlying differences in cultural norms affect the ethical status of conversational AI. Secondly, researchers may want to study the configurations which incorporate the implications of our work to algorithm design and AI implementation. Thirdly, it would be useful to design mechanisms that may enforce AI and other negotiation participants to be ethical.

As AI chatbots become more sophisticated and more closely mimic both ``bad" and ``good" human behaviors, researchers and practitioners are presented with an opportunity to re-evaluate conversational norms that are considered to be ethical vs. unethical. We hope our work inspires future streams of research focused on how the ethics of lies in negotiation vary between human-to-human and AI-to-human interactions, as well as cross-culturally.

\section*{Appendix: But can AI lie?}
Condition (iii) in the ethics of lying in negotiation above is from a trust-based theory of lying \citep{fried1978right}. To make sense of the human-oriented verbs, a fitting theory of mind is needed \citep{isaac2017white}. In our working definitions, the liar must have the capacity to \textit{invite or assure} a human to trust statements. To invite or assure is in part an intentional act. It seems far-fetched to believe that AI, for instance, a conversational agent, can \textit{intentionally} do something. It depends upon how we define intention, of course. In this paper, we do not defend or further develop a particular theory of intention. Instead, we want to note that several philosophers have developed non-human-animal theories of ``deceptive signalling'' and non-human animals, especially domesticated animals, have been considered to be a benchmark in discussing AI's liability issues \citep{asaro201111, bendel2016considerations}. 

Biologists widely study animals' deceptive signals. For instance, wild tufted capuchin monkeys use alarm calls in a functionally deceptive manner to usurp food resources; some male fireflies mimic the mating signals of female fireflies to lure males and destroy them.

Philosopher Skyrm's analysis is a standard non-human theory of deceptive signals, on which others critically further develop their theories. Skyrm writes as follows: 

\begin{quote}
    ``...if receipt of a signal moves probabilities of states it contains information about the state. It moves the probability of a state in the wrong direction--either by diminishing the probability of the state in which it is sent, or raising the probability of a state other than the one in which it is sent--then it is misleading information, or \textit{misinformation}. If misinformation is sent systematically and benefits the sender at the expense of the receiver, we will not shrink from following the biological literature in calling it deception'' \citep[p. 80]{skyrms2010signals}.
\end{quote}

By ``systematically,'' Skyrm means that sending a deceptive signal is distinct from mistakenly sending a signal that is not true, because through evolution and natural section animals are disposed to sending deceptive signals. Likewise, machine learning agents can systematically send a lying report, since machine learning algorithms are trained to consistently maximize a given goal. 

If we use the non-human theory perspective, we can redefine our frameworks, as follows:

\begin{quote}
    Let ``{\em x} sends a lying report/signal to {\em y}'' := There is a proposition {\em P} such that (i) P is not true and (ii) {\em x} systematically sends {\em P} to {\em y}.

    Let ``{\em x} sends a lying report to {\em y} in an ethically impermissible manner'' := There is a proposition {\em P} such that (i) {\em P} is not true, (ii) {\em x} systematically sends {\em P} to {\em y}, and (iii) {\em x} systematically sends the assuring/invitational signal for {\em y} to believe that {\em x} does not lie to {\em y} but it lies.
    
    Let ``{\em x} sends a lying report to {\em y} in an ethically permissible manner'' := There is a proposition {\em P} such that (i)  {\em P} is not true and (ii) {\em x} systematically sends {\em P} to {\em y}, and (iii) {\em x} does not send the assuring/invitational signal for {\em y} to believe that {\em x} does not lie to y. 
\end{quote}

The non-human account, which does not need the term ``intention'' is semantically different from, but substantively the same as the accounts in the main text. To clarify that, the chosen definition of lying above does not use the term ``intention'' and minimally relies on the mental states of the AI \citep{chisholm1977intent}.

\singlespacing

\end{document}